\documentclass{article}
\usepackage{PRIMEarxiv}

\usepackage[utf8]{inputenc} 
\usepackage[T1]{fontenc}    
\usepackage[hidelinks]{hyperref}       
\usepackage{url}            
\usepackage{booktabs}       
\usepackage{amsfonts}       
\usepackage{nicefrac}       
\usepackage{microtype}      
\usepackage{lipsum}
\usepackage{fancyhdr}       
\usepackage{graphicx}       
\usepackage{multirow}
\usepackage[numbers,sort&compress]{natbib}
\graphicspath{{media/}}     
\usepackage{enumitem}

\title{Coding with AI: From a Reflection on Industrial Practices to Future Computer Science and Software Engineering Education
}

\author{
  Hung-Fu Chang \\
  R. B. Annis School of Engineering \\
  University of Indianapolis \\
  Indianapolis\\
  \texttt{hchang@uindy.edu} \\
   \And
  Mohammad Shokrolah Shirazi  \\
  E. S. Witchger School of Engineering \\
  Marian University \\
  Indianapolis\\
  \texttt{mshokrolahshirazi@marian.edu} \\
     \And
  Lizhou Cao  \\
  University of Maryland Eastern Shore \\
  Princess Anne\\
  \texttt{lcao@umes.edu} \\
     \And
  Supannika Koolmanojwong Mobasser  \\
  The Boehm Center for Systems and Software Engineering \\
  Los Angeles\\
  \texttt{supannika@gmail.com } \\
}

\begin{document}
\maketitle

\begin{abstract}
Recent advances in large language models (LLMs) have introduced new paradigms in software development, including vibe coding, AI-assisted coding, and agentic coding, fundamentally reshaping how software is designed, implemented, and maintained. Prior research has primarily examined AI-based coding at the individual level or in educational settings, leaving industrial practitioners’ perspectives underexplored. This paper addresses this gap by investigating how LLM coding tools are used in professional practice, the associated concerns and risks, and the resulting transformations in development workflows, with particular attention to implications for computing education. We conducted a qualitative analysis of 57 curated YouTube videos published between late 2024 and 2025, capturing reflections and experiences shared by practitioners. Following a filtering and quality assessment process, the selected sources were analyzed to compare LLM-based and traditional programming, identify emerging risks, and characterize evolving workflows. Our findings reveal definitions of AI-based coding practices, notable productivity gains, and lowered barriers to entry. Practitioners also report a shift in development bottlenecks toward code review and concerns regarding code quality, maintainability, security vulnerabilities, ethical issues, erosion of foundational problem-solving skills, and insufficient preparation of entry-level engineers. Building on these insights, we discuss implications for computer science and software engineering education and argue for curricular shifts toward problem-solving, architectural thinking, code review, and early project-based learning that integrates LLM tools. This study offers an industry-grounded perspective on AI-based coding and provides guidance for aligning educational practices with rapidly evolving professional realities.
\end{abstract}

\keywords{Vibe Coding; Agentic Coding; AI-based Coding; LLM}

\section{Introduction}
Recent advances in large language models (LLMs) have fundamentally transformed software development through their powerful code generation capabilities. Unlike traditional programming, which requires developers to write code line by line, this new approach uses a natural language input, called a prompt, to describe a problem or task to automatically produce the desired code. This "text-to-code" process has led to "vibe coding," a term introduced by Andrej Karpathy in early 2025 to capture the practice of expressing one's intent in natural language and letting AI generate the program \cite{karpathy_2025}. In a broader sense, vibe coding can be understood as a new mode of software creation that minimizes direct programming effort, enabling users to concentrate on the desired functionality or final product without necessarily reviewing, testing, or even fully understanding the generated code \cite{sapkota_2025_vibe, sarkar_2025_vibe}. This new development method draws many users who have not been through programming training to participate software development with LLM tools \cite{gadde_2025_democratizing}.

Building upon vibe coding, agentic coding extends this development method by delegating not only code generation but also substantial development activities to autonomous AI agents \cite{sapkota_2025_vibe, wang_2025_ai}. In agentic coding, AI systems act on behalf of the user to plan, execute, and coordinate multiple tasks, such as setting up development environments, modifying codebases, running tests, and managing deployment workflows, often with minimal human intervention. While vibe coding emphasizes prompt-driven, conversational code generation, agentic coding operationalizes these interactions into goal-directed, multi-step processes, effectively transforming natural language intent into coordinated software engineering actions under human supervision.

This paradigm shift has significant implications for both the software engineering community and industry, particularly through the integration of LLM-enhanced coding tools into mainstream workflows. These tools replace much of the traditional hand-crafted coding process with dynamic human–AI collaboration, generating large portions of code rapidly without requiring formal specification. AI-generated code is no longer a supplement but a primary driver of software development. Furthermore, the tool further moves from merely code generation to agentic coding, which is delegating substantial coding tasks to autonomous AI agents within the development environment to execute multiple development tasks. As a result, a recent industry report on Silicon Valley startups (e.g., Y Combinator cohorts) noted that 25\% of teams had as much as 95\% of their codebase generated via LLM-enhanced coding tools \cite{mehta_2025_a}. Likewise, Microsoft CEO Satya Nadella revealed that 20\% to 30\% of the company’s repository code was written by AI \cite{zeff_2025_microsoft}. The implication of this phenomenon is clear. On one hand, vibe coding accelerates development timelines and lowers the barrier to entry by allowing even non-experts to create functional applications through descriptive prompts and speed up the development through generating repetitive or tedious code snippets. On the other hand, LLM coding tools that embedded autonomous agents further change the way software developers do their jobs. At the same time, its growing adoption has fueled provocative debates, with some in industry and academia ambitiously suggesting that if natural language intent can fully replace computational thinking and syntax, traditional programming may be facing obsolescence.

The widespread adoption of vibe and agentic coding on the software community have prompted engineers, practitioners, and researchers to investigate how this new approach reshapes existing development practices. However, existing research focuses on the influence of vibe coding, agentic coding, and LLM tools at the individual level, their effects on team dynamics and collaborative workflows, often being seen in industry, remain largely unexplored.  In addition, there is still no research that systematically examined AI usage in coding from the perspective of industry. 
To fill up this gap, we propose a systematic approach by investigating industrial reports, discussions, and interviews about using AI in coding by collecting industrial’s opinions from YouTube. Its videos have become a valuable resource for studying industrial practices and experiences, since the platform has become widely used for sharing knowledge and teaching real-world practices. We expect to answer several questions that continue to produce debate within the research community: How should vibe or agentic coding be defined? How do they differ from traditional code generation? Collectively, these questions point to a broader, overarching inquiry: What new processes and methodologies will emerge in software development, and how should education adapt to prepare developers for these changes? Furthermore, we summarize current practices and identified key differences between traditional coding methods and the emerging vibe coding approach in software development processes. We expected to highlight the new changes needed for education as well as potential adjustments to software development lifecycle through the understanding of these working practice changes.

\section{Methodology}
\subsection{Research Questions and Further Investigations}
In this paper, we review and analyze practitioners’ perspectives to address the following research questions and to further discuss how software developers should prepare for emerging changes in software development.

\textbf{RQ1}: What types of AI usage exist in programming, and how do they differ from traditional software development approaches?

\textbf{RQ2}: What concerns and risks does the adoption of AI tools introduce into the software development process?

\textbf{RQ3}: How does the integration of AI tools reshape software development workflows, and what new skills and competencies are required of software developers?

For RQ1, we synthesize practitioners’ opinions on the role of AI tool in programming, which helps clarify new terminology and emerging concepts related to AI usage in software development. We also draw on academic views to compare programing using AI tools with traditional software development. RQ2 examines the key concerns and risks arising from the use of AI tools in software development. RQ3 investigates how software development workflows are evolving and identifies new skills and competencies needed for developers to adapt these changes. Finally, we integrate findings across all three research questions to discuss their broader implications and to highlight potential directions for software engineering education.

\subsection{Approach}
\begin{figure}[h!]
  \centering
  \includegraphics[width=0.5\columnwidth]{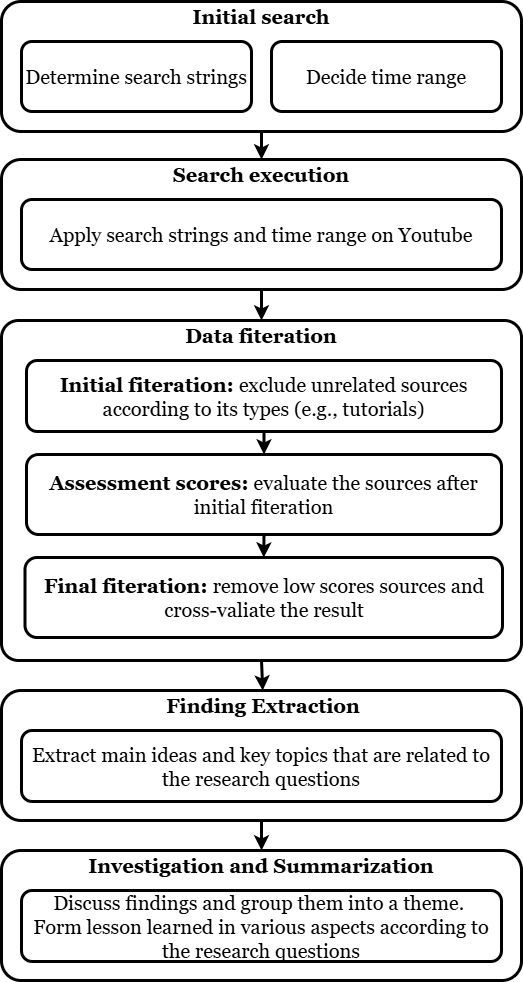}
  \caption{Flowchart of Our Data Collection Approach}
  \label{fig:Flowchart_of_Approach}
\end{figure}

Our research approach consists of five major stages, as illustrated in \autoref{fig:Flowchart_of_Approach}. These stages outline the systematic process used to identify, filter, and analyze relevant sources, as well as to extract and synthesize key findings related to the research questions. The following sections describe each stage in detail.

\subsubsection{Search String and Time Range}
Data were collected from YouTube videos published between December 15, 2024, and October 25, 2025, as Google Search Trends indicate that interest in AI coding began to rise in late December 2024 and vibe coding emerged as a distinct trend around mid-February 2025. YouTube was selected as the data source because it captures practitioner discussions and provides diverse perspectives on contemporary software development practices. The search strings used for video retrieval were “vibe coding,” “AI coding,” “AI programming,” “AI in coding,” and “AI in software development.”

\subsubsection{Search String and Time Range}
The initial search results were filtered by the lead author based on the type of content. Our focus was on in-depth experiences and reflections regarding the use of AI in software development shared by experienced software professionals. Consequently, videos such as tutorials, tool comparisons, or demonstrations by non-developers were excluded. Specifically, two categories of videos were removed: (1) videos teaching “vibe coding,” demonstrating vibe coding tools, or comparing software tools, and (2) videos presenting personal opinions from individuals with little or no software development experience. Ultimately, three types of sources were retained (see \autoref{tab:source_type_of_data}). 

\begin{table}
 \caption{Source types of final data collection}
  \centering
  \renewcommand{\arraystretch}{1.35}
  \begin{tabular}{lll}
    \toprule
    Video Type     & Description & Number of Videos \\
    \midrule
    Opinion/ Experience       & \parbox[t]{3.5in}{Personal opinions are solely delivered in this type of video. This might also include some critiques or demonstration. Personal experiences are solely delivered in this type of video. It could also provide guide, best practice, case study, or analysis for some aspects related to AI-based coding.} & 21   \\
    Presentation         & \parbox[t]{3.5in}{This is a recording about public presentations or talks in a conference.} & 9   \\
    Discussion  & \parbox[t]{3.5in}{This type of video includes discussions among multiple people or one-to-one interviews. The video often involves many in-depth questions and answers between participants.} & 27  \\
    \bottomrule
  \end{tabular}
  \label{tab:source_type_of_data}
\end{table}

Inspired by the guideline suggested by Garousi et al. \cite{garousi_2019_guidelines}, the data after filtering were then assessed by the authors by scoring from 0 to 5 across the following five quality dimensions.

    \begin{itemize}
        \item \textbf{Authority of the producer:} Credibility of the author, such as a well-known individual or affiliation with a reputable organization.
        \item \textbf{Evidence:} Inclusion of examples, rationale, or empirical support.
        \item \textbf{Objectivity:} Presentation or statement of ideas in a neutral and balanced manner. 
        \item \textbf{Content Type:} Reflection on or explanation of industrial practices and experiences.
        \item \textbf{Novelty:} Contribution of unique insights or additions to existing research.
    \end{itemize}

Videos that scored above 18 points were retained for further cross-validation, which was conducted by coauthors to enhance objectivity. After two rounds of assessments, 57 videos were finally selected from the initial 224 search results. This final dataset forms the basis for analyzing the research questions and provides insight into trends, adoption practices, and concerns within the software engineering community.

\subsubsection{Finding Extraction}

In this extraction process, the filtered sources were distributed among the authors for systematic reviews and ideas extraction. The extracted ideas were subsequently discussed, validated, and synthesized. Finally, the topics were organized into thematic groups corresponding to research questions. \autoref{tab:theme_topic} presents the themes derived from the analyzed videos. 

\begin{table}
\caption{Themes under each research question}
\begin{center}
\begin{tabular}{|c|p{5.5in}|} 
    \hline
      & Theme \\ 
    \hline
    RQ1 & 
        \begin{minipage}[t]{\linewidth}
        \begin{itemize}[nosep,after=\strut]
            \item Definition and philosophy of Vide Coding
            \item Definition and philosophy of Agentic Coding
            \item AI assisted development
            \item A new way of code generation - deterministic vs. stochastic
            \item Shifted developer's role
        \end{itemize} 
        \end{minipage} \\
    \hline
    RQ2 & 
        \begin{minipage}[t]{\linewidth}
        \begin{itemize}[nosep,after=\strut]
            \item Security risks and vulnerabilities
            \item Concerns about removing the training ground necessary for juniors to become experts
            \item Technical debt and maintainability
            \item Bypassing traditional safety pipelines
            \item Code review as a bottleneck
            \item Ethical concerns
            \item Impact on developer’s job satisfaction
        \end{itemize} 
        \end{minipage} \\
    \hline
    RQ3 & 
        \begin{minipage}[t]{\linewidth}
        \begin{itemize}[nosep,after=\strut]
            \item Rapid prototyping and experimentation
            \item Code review as the new primary skill
            \item Faster and more frequent iterations
            \item Documentation formats
        \end{itemize} 
        \end{minipage} \\
    \hline
\end{tabular}
\end{center}
\label{tab:theme_topic}
\end{table}

The findings from these sources provide a comprehensive understanding of the impacts of AI-based coding on industrial software development and inform further analysis regarding needs for education in this context.

\section{Result}
\subsection{Findings for RQ1}
Five themes are identified under RQ1, describing various types of AI usages in software development and comparing them with traditional coding practices. The following sections summarize these findings.

\subsubsection{AI Usage in Software Development}
The findings highlight how AI is used in software development, contrasted with traditional coding practices, and include considerations regarding enterprise adoption of AI tools. Industrial perspectives are discussed.
The use of AI in software development is not entirely new. For instance, in requirement analysis, Natural Language Processing (NLP) and Machine Learning (ML) have been applied to requirement elicitation and specification classification \cite{liu_2022_artificial}. In software testing, ML and AI techniques have been employed for test case generation, test case prioritization, and defect detection \cite{islam_2023_artificial}. In this study, the AI technologies under investigation are primarily Large Language Model (LLM) tools.

\paragraph{Definition of Vibe Coding}
Vibe coding is characterized by an intuitive, “surrender to the flow” mindset—sometimes described as “forgetting that code even exists.” Users provide high-level instructions in natural language prompts and expect functional output without directly engaging in code. This approach largely bypasses traditional software engineering practices. Example quotes from our collected data include:

        \begin{itemize}[nosep,after=\strut]
            \item \textit{“Worth giving it a new designation of software 3.0 and basically your prompts are now programs that program the LLM”}
            \item \textit{“I think that the whole idea of kind of vibe coding or coding without really looking at the code and understanding it”}
            \item \textit{“Maybe you've seen a lot of GitHub code is not just like code anymore there's a bunch of like English interspersed with code”}
            \item \textit{“Forget that code even exists how Andrej Kaparthy created that vibe coding idea ... that's the YOLO vibe coding and not meant to ship to production it's about speed instant gratification there's creativity in there and it's really about this fast learning hopefully”}
        \end{itemize}

Vibe coding involves an "iterative goal-satisfaction cycle," where users formulate high-level goals and iteratively prompt the AI model, often through voice-to-text tools, to generate and refine code. The user's role is not to write code, but to orchestrate its creation and rapid evaluation in an almost exploratory manner. This concept, referred to as "material disengagement" from the code itself, emphasizes continuous collaboration with LLM without full coding involvement. 

\paragraph{Definition of Agentic Coding}
Industrial practitioners describe agentic coding as delegating coding tasks to autonomous AI agents, similar to collaborating with multiple engineers. Representative quotes can be described as follows.

        \begin{itemize}
            \item \textit{“The idea of like an agent provisioning a development environment installing packages provisioning databases deploying for you”}
            \item \textit{“Human guiding the overall process essentially working as an orchestrators of all the task and AI programs call them agents understand what human wants to achieve and they make it happen by translating the human intent into action and typically it will be a multi-agent system including things like uh specification agents architecture agents uh coding agent testing agent deployment agent so on and so forth”}
            \item \textit{“Really goes from the idea of like coding with an assistant to really delegating work to an agent”}
            \item \textit{“Actually I want you to go do this task I want you to go do this task as you're working like it can come back an hour later”}
        \end{itemize} 

Developers divide a task or a system feature into smaller units, delegate to the agents, and synthesize outputs after testing. That is, agentic coding can be defined as a paradigm in AI-assisted programming where developers delegate substantial coding tasks to autonomous AI agents that operate within the development environment. Agentic coding tools (e.g., Anthropic’s Claude Code, OpenAI’s Codex) can autonomously read and modify codebases, execute tests, run shell commands, and manage CI/CD workflows, subject to human oversight. 

\begin{figure}[hbt!]
  \centering
  \includegraphics[width=0.9\columnwidth]{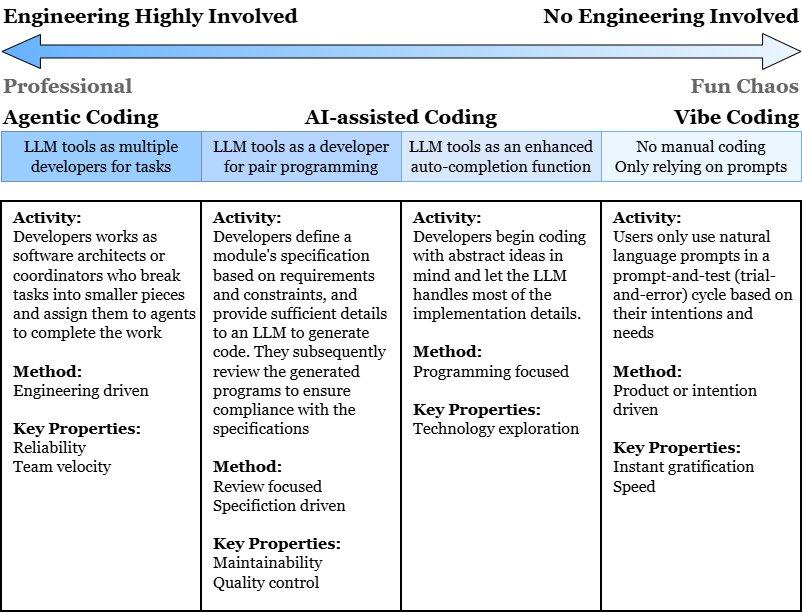}
  \caption{Spectrum of AI-based Coding}
  \label{fig:spectrum_of_ai_coding}
\end{figure}

\paragraph{Summary of AI-based Coding Definitions}
Drawing on practitioners’ perspectives (mainly from "Vibe Coding at Scale: Customizing AI Assistants for Enterprise Environments" presented), we summarize AI-based coding practices along two key dimensions: the level of LLM involvement and the degree of human engineering effort, as reflected in coding behaviors. Based on these dimensions, three distinct coding paradigms can be identified: vibe coding, AI-assisted coding, and agentic coding. Although all three paradigms rely on LLM technologies, they differ fundamentally in their underlying philosophy, development processes, and implications for software engineering practice.

\autoref{fig:spectrum_of_ai_coding} illustrates the spectrum of activities derived from our synthesis of practitioners’ opinions, organized by the level of human engineering involvement. At one extreme is vibe coding, which is characterized by minimal human intervention and a prompt-driven, exploratory approach to software creation. At the opposite extreme is agentic coding, in which autonomous AI agents carry out substantial development tasks—such as planning, implementation, testing, and deployment—under human orchestration. AI-assisted coding occupies the intermediate position, where human developers remain actively engaged in design and implementation while using AI tools to augment productivity rather than fully delegate coding responsibilities.

\subsubsection{AI-based and Traditional Coding}
\paragraph{Comparison of AI-based and Traditional Coding}
The distinction between traditional coding and AI-based coding is often first observed in tool usage. Traditional development environments typically provide syntax highlighting and statement-level auto-completion, whereas LLM coding tools are capable of generating code snippets or even entire programs. However, the differences extend beyond tooling and encompass task scope, user roles, developer responsibilities, and levels of software engineering involvement. The lessons learned from our analysis are summarized in \autoref{tab:comparison_trad_and_ai}.

Traditional coding supports tasks ranging from simple functions to highly complex, tightly integrated systems. Developers take full responsibility for system design, implementation, and integration, relying on their ability to reason about system-wide architecture and module interactions. Programming in this paradigm requires explicit specification of implementation details and deep technical expertise.

In contrast, AI-based coding spans a broader range of user expertise, from individuals with no programming background to experienced software engineers. In vibe coding, users interact with AI almost exclusively through natural language prompts and conversational feedback, without engaging in systematic decomposition, modular integration, or architectural reasoning. As a result, vibe coding is poorly suited for developing complex systems with multiple interdependent components.

In AI-assisted coding and agentic coding, developers maintain a central role. They decompose large or abstract problems into manageable modules, validate AI-generated snippets, and integrate them into the overall system. These approaches still require strong software engineering knowledge, as humans remain responsible for architectural decisions and quality assurance, while AI primarily serves as a productivity enhancer. As a result, AI-based coding tools are often adopted by users focused on rapid prototyping and product exploration, such as product managers, prototype builders, and less experienced software engineers.

\begin{table}
 \caption{Comparison between traditional and AI-based coding}
  \centering
  \renewcommand{\arraystretch}{1.35}
  \begin{tabular}{lll}
    \toprule
       & Traditional & \parbox[t]{2.5in}{AI-based (Vibe, AI-assisted, and Agentic)} \\
    \midrule
    \parbox[t]{1.75in}{Scope of coding task} & \parbox[t]{1.75in}{From functional to system level} & \parbox[t]{2.5in}{Mostly functional level; Some from functional to system level} \\
    \parbox[t]{1.75in}{Coding knowledge and \par experience level} &  \parbox[t]{1.75in}{High}  &  \parbox[t]{2.5in}{Zero to high} \\
    \parbox[t]{1.75in}{Potential users}  & \parbox[t]{1.75in}{Software engineer} & \parbox[t]{2.5in}{Inexperienced user, customer engineer, software engineer}  \\
    \parbox[t]{1.75in}{Developer role}  & \parbox[t]{1.75in}{Code author} & \parbox[t]{2.5in}{Specification writer, reviewer, editor of AI suggestions}  \\
    \parbox[t]{1.75in}{Level of software engineering practice involvement}  & \parbox[t]{1.75in}{High} & \parbox[t]{2.5in}{Zero to high}  \\
    \parbox[t]{1.75in}{\raggedright Tool assistance involvement}  & \parbox[t]{1.75in}{\raggedright Editing, syntax highlighting, auto completion} & \parbox[t]{2.5in}{\raggedright Code generation}  \\
    \parbox[t]{1.75in}{Tool interaction}  & \parbox[t]{1.75in}{Code editing} & \parbox[t]{2.5in}{Prompt and code editing} \\
    \bottomrule
  \end{tabular}
  \label{tab:comparison_trad_and_ai}
\end{table}

\paragraph{Code Generation: Traditional vs AI-Based Approaches}
Traditional code generation relies on formal or semi-formal models and specifications, such as UML or Z \cite{brambilla_2022_modeldriven}. Requirements are translated into precise logical representations, which are then transformed into source code through deterministic model-to-code processes. Because each step is explicitly defined, this approach provides strong guarantees of correctness and traceability from requirements to implementation \cite{ramesh_2001_toward}. However, it demands substantial expertise in formal modeling and entails a steep learning curve.

However, code generation using Large Language Models (LLMs) shifts interaction toward natural language. Developers express intent through prompts, examples, and dialog, and the AI generates code. This process is probabilistic; that is, slight variations in prompt phrasing or context can lead to different outputs, a phenomenon widely recognized by industrial practitioners:

        \begin{itemize}
            \item \textit{“It tries to complete the next word the next token as we call it and the way it happens it generates a lot of candidates”}
            \item \textit{“The same input would result in different outputs”}
        \end{itemize} 

Correctness is therefore validated primarily through testing rather than guaranteed by construction \cite{liu_2023_is} and maintaining traceability from requirements to code can be challenging. \autoref{tab:comparison_trad_and_llm_codegen} summarizes the comparison between these two approaches.

\begin{table}
 \caption{Comparison between traditional and LLM code generation}
  \centering
  \renewcommand{\arraystretch}{1.35}
  \begin{tabular}{lll}
    \toprule
       & \parbox[t]{2.5in}{Traditional Code Generation} & \parbox[t]{2.5in}{Code Generation via LLMs} \\
    \midrule
    \parbox[t]{1in}{Input} & \parbox[t]{2.5in}{Formal / semi-formal (ex: UML, Z, or etc.)} & \parbox[t]{2.5in}{Natural language + examples + dialog (prompt + history)} \\
    \parbox[t]{1in}{Determinism} &  \parbox[t]{2.5in}{Deterministic model-to-code transformations, reproducible builds.}  &  \parbox[t]{2.5in}{Outputs vary with prompt phrasing, model state, and sampling.} \\
    \parbox[t]{1in}{Correctness guarantees}  & \parbox[t]{2.5in}{Supports proofs, refinement, model checking to correct-by-construction code.} & \parbox[t]{2.5in}{Relies on empirical validation (tests, review). Susceptible to hallucinations and subtle bugs.}  \\
    \parbox[t]{1in}{\raggedright Traceability}  & \parbox[t]{2.5in}{\raggedright Requirements - models - code mapping explicit; strong compliance/audit support.} & \parbox[t]{2.5in}{\raggedright Weak unless prompt/versioning discipline added; chat logs can help but are ad-hoc.}  \\
    \parbox[t]{1in}{Maintainability}  & \parbox[t]{2.5in}{Enforced architecture and consistent patterns via templates.} & \parbox[t]{2.5in}{Traceability is ad hoc (prompt logs). Weak support without added governance.}  \\
    \parbox[t]{1in}{Expertise}  & \parbox[t]{2.5in}{Requires modeling/formal methods expertise. Steep learning curve.} & \parbox[t]{2.5in}{Requires prompt design, critical review, orchestration. Lower entry barrier for domain experts.}  \\
    \bottomrule
  \end{tabular}
  \label{tab:comparison_trad_and_llm_codegen}
\end{table}

\subsubsection{Using LLM Tools to Aid Software Development}
One obvious distinguishing feature of AI-assisted development is its ability to leverage LLMs’ learning, reasoning, and rapid code generation capabilities. Industrial practitioners identified three strategic practices for effectively integrating AI into software development.
First, developers must establish a strong contextual foundation, including familiarizing the AI with the codebase, coding standards, project-specific rules, and architectural patterns. This enables AI tools to explore components, trace data flows, and explain system architecture, particularly in unfamiliar codebases. Companies are encouraged to define global coding standards, project-specific rules, and provide contextual artifacts such as code snippets or UI screenshots.
Second, LLMs can support rapid prototyping and idea exploration, especially during early-stage development. This capability allows teams to quickly validate ideas without committing to production-quality code:
        \begin{itemize}
            \item \textit{“We're just doing this to get feedback. It just needs to be a prototype and that's really where some of these vibe coding tools would do”}
        \end{itemize} 
Third, LLM tools enable participation from non-technical stakeholders by lowering the barrier to experimentation. Non-technical team members can explore design alternatives, conduct product research, and interact with APIs or documentation through an LLM tool:
        \begin{itemize}
            \item \textit{“I can introduce more non- STEM disciplines into software development cycles.”}
        \end{itemize} 

\subsection{Findings for RQ2}
Seven themes were identified under RQ2, capturing industry concerns related to the adoption of AI in coding. These themes are discussed in the following sections.

\subsubsection{Concerns in AI-based Coding}
The adoption of LLM in software development introduces several critical concerns. LLM-generated code may be inconsistent, faulty, or risky, negatively affecting code quality, maintainability, and security. Addressing these issues often requires additional development effort for debugging, refactoring, or structural correction. Moreover, experienced developers express concern about skill erosion and the loss of enjoyment traditionally associated with programming. Delegating decision-making to LLM also raises questions about accountability, responsibility, and trust.

\paragraph{Loss of problem-solving enjoyment}
Several practitioners highlighted the intrinsic satisfaction derived from manual problem-solving and hands-on coding:

        \begin{itemize}
            \item \textit{“I want to solve a problem that requires no internet that requires no starting of anything. I want to just be able to just solve the thing quickly and it actually feels really nice.”}
            \item \textit{“AI coding is undermining the enjoyment of programming”}
        \end{itemize} 

In vibe coding, the intellectual engagement associated with writing, experimenting, and refining code is regarded as the most rewarding aspects of software development for many programmers. This now is replaced by automatically code generation. AI tools may just leave programmers with the final product but without valuable learning experience and creative fulfillment that comes from actively wrestling with code. 

\subsubsection{Code quality and increased efforts}
Production-grade software demands good quality by using predefined workflow to test and verify code. Practitioners emphasize the need for making incremental progress in small and controlled steps, verifying changes continuously to ensure correctness, and keeping modular system, when working with AI-generated code:

        \begin{itemize}
            \item \textit{“When working with an AI we must be able to verify after every small change”}
        \end{itemize} 

Some professionals view vibe coding as fundamentally incompatible with engineering discipline, to poorly structured and unmaintainable code (e.g., spaghetti code). This kind of messy, broken and embarrassing program fails to meet professional standards.  AI-written code frequently contains bugs or vulnerabilities, suffering from quality, maintainability, and security issues. Therefore, necessitating careful human review or major adjustment on the code structure or bug fix may increase the development efforts. 

\subsubsection{Security and data privacy}
Security concerns largely occupy industrial discussions on AI generated code. One observation is that LLM tools prioritize generating plausible outputs to its users quickly so it may overlook security best practices. Empirical evaluations suggest that only a small proportion (i.e., around 35\%) of AI-generated backend code is both secure and correct. While some models perform adequately well on SQL injection avoidance or cryptographic algorithm, they frequently mishandle cross-site scripting and log injection vulnerabilities.

Prior research showed that even "security-focused prompts” fail to eliminate vulnerabilities \cite{denny_2024_computing}. A famous example in industry is the Replit AI catastrophe in July 2025, which occurred during a vibe-coding session. Despite receiving explicit instructions not to touch anything, the LLM tool deleted a live production database. The incident illustrates that AI-generated code can bypass standard security checks, leading to unseen vulnerabilities.  Similar failures involving private data leaks in a dating app expose critical systemic flaws in the vibe coding mindset. The reliance on an AI's internal "thinking" without an external verification process is fundamentally "untrustworthy behavior". The true cost of disengagement is not just a messy codebase but a direct threat to user data, operational stability, and brand reputation. This is particularly concerning for novice programmers who may lack the knowledge to provide the necessary human oversight and may be more likely to believe their code is secure, even when it is not.  The opaque, black-box nature of LLMs makes it challenging to ascertain the security posture of the generated code.

One solution for companies to enhance their security is to build an agent to add enterprise-grade security check and governance:

        \begin{itemize}
            \item \textit{“The launch of agent aims to solve the challenge of using vibe coding in production by adding enterprise-grade security and governance”}
        \end{itemize} 
        
Such agents can perform automated security checks and audits before deployment, mitigating some of the associated risks. 

\subsubsection{Ethical concerns}
AI-based coding raises significant ethical and legal concerns, particularly regarding Intellectual Property (IP). Key question include ownership of AI-generated code and whether creators should be compensated for the data used to train these models \cite{chesterman_2024_good}. Companies express concern over potential legal issues when someone copies the generated code without modifications. Many organizations choose to train models exclusively on proprietary data:

        \begin{itemize}
            \item \textit{“We need to train our model this was the Salesforce code generation model and the enterprise wants to use their own IP.”}
            \item \textit{“I think that there's already been legal precedence if somebody takes code from AI, it's not theirs. If somebody takes code from AI and modifies it it's theirs… I think by current legal precedent they're not the owner of that code that would be a heck of a loophole”}
        \end{itemize} 

These concerns underscore the need for clearer legal frameworks and ethical guidelines governing AI-based coding.

\subsection{Findings for RQ3}
\subsubsection{Boosted Development Tasks and Bottleneck Shifting}
Many sources indicate that LLM tools significantly accelerate software development tasks, thereby potentially shifting the primary bottleneck in the development life-cycle from code writing to review. 

Industrial practitioners observe that as AI generates larger volumes of code more rapidly, the effort required to read, understand, and validate this code increases substantially. Because of increasing importance of reading and debugging codes, companies think code review should be a critical skill to be checked in the hiring interviews. 

The following quotes illustrate these industry observations:

        \begin{itemize}
            \item \textit{“Write three to four times more code but submit fewer and larger. Human reviewers might write less code manually but spend much more time shifting”}
            \item \textit{“The next thing is code review is by far the most important skill, I think. We probably should have been interviewing for code review”}
            \item \textit{“I also think that we're going to be seeing new bottlenecks in the software development life cycle specifically around both code and application review”}
            \item \textit{“My opinion like code review as the interview versus actually producing code”}
        \end{itemize}

In addition, organizations can also produce AI agents to automate traditional manual development tasks. The following quote illustrates one possible application of AI in practice. Notably, as discussed in the previous section, the use of AI agents for code auditing also contributes to improved development efficiency and quality.

        \begin{itemize}
            \item \textit{“Efficiency is maximized by using multiple agents simultaneously and streamlining the review process through automated testing, security scanning, and CI/CD review apps.”}
        \end{itemize} 

\subsubsection{Personal skill erosion and threat to beginners}
A major concern associated with vibe coding is the potential erosion of developers’ technical skills due to reduced hands-on coding practice. Over time, heavy reliance on AI may weaken problem-solving abilities and diminish confidence in reasoning about software behavior. As stated in many interview and discussion video sources, practitioners emphasize that an engineer’s value lies in critical thinking and problem-solving skills rather than familiarity with specific tools.

This concern poses a particular threat to beginners who are trying to cultivate foundational skills. Although LLM tools can bring immediate satisfaction by enabling rapid creation of full-stack applications with minimal coding effort, such acceleration carries significant risks. LLM code generation process can prevent beginners from developing a deep understanding of why code works. 

\begin{figure}[hbt!]
  \centering
  \includegraphics[width=0.5\columnwidth]{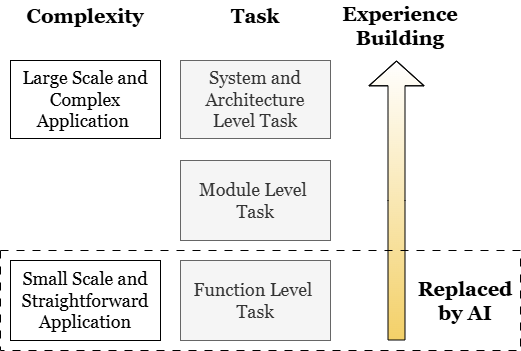}
  \caption{The missing part of the experience growth for an engineer}
  \label{fig:missing_part_of_growth}
\end{figure}

\autoref{fig:missing_part_of_growth} illustrates the traditional trajectory of experience accumulation in software companies. Engineers typically build expertise by implementing function-level components and small-scale applications before progressing to system-level jobs. Industrial practitioners express concern that this pathway is being disrupted:

        \begin{itemize}
            \item \textit{“Need enough knowledge to judge what's good versus bad and you only become good with enough practice”}
        \end{itemize} 

As AI increasingly replaces function-level tasks, opportunities for experiential learning diminish. This phenomenon creates what we term a “paradox of skill development”. Some practitioners warn that stopping hiring entry level engineers raises a basic question - \textit{who is going to train inexperienced engineers?} Moreover, many discussions state that, in AI era, the critical skill is not about producing answers but knowing how to ask the right question. Therefore, without enough accumulation on basic skills and foundational problem-solving, developers may struggle to ask the right question?

\subsubsection{Role Transformation in Software Development}
AI-based coding introduces significant role changes for software engineers. Instead of serving primarily as code producers, engineers increasingly act as task facilitators and idea creators. Practitioners often describe this shift metaphorically as role becoming a conductor:

        \begin{itemize}
            \item \textit{“Our skills as experienced full-stack designers/developers are now to direct AI to carry out our design.”}
            \item \textit{“The role is more... if you think of it like a conductor of sorts as opposed to the actual instrument player.”}
        \end{itemize} 
        
Developers interact with large language models (LLMs) through natural language prompts to generate code, shifting their responsibilities from manual programming to guiding, specifying, and refining AI-generated outputs \cite{pajo_2025_vibe}. This transformation represents a fundamental departure from traditional programming workflows, requiring developers to manage multiple AI agents similarly to a distributed team.

This shift also redistributes cognitive load. Instead of grappling with syntax, APIs, and repetitive implementation details, developers devote greater attention to domain modeling, architectural decisions, and system integration. In this sense, AI-based coding unifies design, implementation, and validation into an intent-driven process, where natural language becomes the primary medium of software composition rather than an auxiliary form of documentation.

By enabling experts to interact directly with executable representations, AI-generated code reduces the need for developers to act as intermediaries translating abstract ideas into formal syntax \cite{li_2025_a}. Vibe coding, by contrast, aligns cognition with execution through dialogue. Developers articulate their intent, the AI generates a candidate solution, and the human iteratively revises it. This conversational cycle fosters cognitive alignment, where mental models and evolving code remain synchronized in near real time. Prior research suggests that such collaboration can reduce extraneous cognitive load by automating routine tasks, allowing developers to focus on higher-order reasoning and creativity \cite{zvielgirshin_2024_the, a2024_towards}. Ultimately, vibe coding shifts the programmer’s role from memorizing technical details to orchestrating a collaborative process, freeing cognitive resources for creativity, problem framing, and strategic decision-making. 

\subsubsection{New Skillset or New Focus on Existing Skillset}
Effective use of LLM tools in software development depends heavily on the clarity and specificity of user instructions. Practitioners consistently note that vague prompts yield unreliable or incorrect outputs, whereas detailed instructions, such as explicitly defining data structures, APIs, or constraints, significantly improve code quality. This implies that prompt engineering should be extended beyond wording to incorporate contextual information throughout LLM usage. As a result, it is important to consider how to cultivate an environment that supports the construction and maintenance of a rich ecosystem of contexts.

        \begin{itemize}
            \item \textit{“Just say please fix and cursor is not able to fix it and any other IDE is not able to fix it because it doesn't know there is no clear instructions and there is no context”}
            \item \textit{“Prompt engineering is all about tweaking wording phrasing things in a specific way to get a single good answer from the LLM. but context engineering supplying all relevant facts rules documents plans tools So the LLM has a whole ecosystem of context”}
        \end{itemize} 

In addition to the new skillset – prompting, many practitioners argue that the most critical skill is not prompting itself, but the ability to read, review, and debug AI-generated code. Code reading involves discerning quality, design trade-offs, and potential risks rather than merely identifying syntactic correctness:

        \begin{itemize}
            \item \textit{“I do think skills of reading code and debugging are maximum is like you have to have the taste and enough training to know that the LLM is bad stuff or good stuff”}
        \end{itemize} 

\subsubsection{Documentation Format}
An emerging practice in AI-assisted development is adapting documentation formats to be more compatible with LLM training and ingestion. Practitioners emphasize the importance of providing documentation in machine-readable and learnable formats, a notion supported by prior research demonstrating improved learning efficiency \cite{chen_2025_mdeval, he_2024_does}:

        \begin{itemize}
            \item \textit{“Offering documentation in LLM-friendly formats (like .txt or Markdown).”}
            \item \textit{““Documentation should change directives like "click" to executable commands (e.g., curl)”.”}
            \item \textit{“Tools are needed to ingest data into LLM-friendly formats (e.g., changing a GitHub URL to get ingest to concatenate files).”}            
        \end{itemize} 

These practices suggest a broader shift toward documentation designed not only for human consumption but also for AI interpretation, further integrating LLMs into the software development ecosystem.

\section{Further Discussion – Impacts on Computing Education}
Industrial practices and practitioner experiences clearly indicate a fundamental transformation in software development driven by the widespread adoption of LLM-based tools. Drawing on the findings of our investigation, this section discusses key implications of this transformation and articulates lessons learned from industry. Based on these insights, we further infer necessary adaptations to computing education, emphasizing the need for closer alignment between academic preparation and contemporary professional practice (summarized in \autoref{tab:changes_in_education}).

\begin{table}
\caption{Changes in Computer Science and Software Engineering Education in Response to Identified Lessons}
\begin{center}
\begin{tabular}{|c|p{3.5in}|} 
    \hline
     \parbox[t]{2.5in}{Educational Transformation} & Discussions on Lessons Learned from Data Analysis \\ 
    \hline
    \parbox[t]{2.5in}{Individual development process change} & 
        \begin{minipage}[t]{\linewidth}
        \begin{itemize}[nosep,after=\strut]
            \item Personal Skill Erosion and Threat to Beginners.
            \item AI Usage in Software Development.
            \item Using LLM Tools to Aid Software Development.
        \end{itemize} 
        \end{minipage} \\
    \hline
    \parbox[t]{2.5in}{Concentrate on Problem-solving} & 
        \begin{minipage}[t]{\linewidth}
        \begin{itemize}[nosep,after=\strut]
            \item Personal Skill Erosion and Threat to Beginners.
            \item Loss of problem-solving enjoyments.
            \item Role Transformation in Software Development.
        \end{itemize} 
        \end{minipage} \\
    \hline
    \parbox[t]{2.5in}{Focus on Architectural Thinking and Specification-Driven Development} & 
        \begin{minipage}[t]{\linewidth}
        \begin{itemize}[nosep,after=\strut]
            \item Role Transformation in Software Development
        \end{itemize} 
        \end{minipage} \\
    \hline
    \parbox[t]{2.5in}{Introducing Project-based Learning Early} & 
        \begin{minipage}[t]{\linewidth}
        \begin{itemize}[nosep,after=\strut]
            \item Security and data privacy
            \item Boosted Development Tasks and Bottleneck Shifting
            \item New Skillset or New Focus on Existing Skillset
            \item Documentation Format
        \end{itemize} 
        \end{minipage} \\
    \hline
\end{tabular}
\end{center}
\label{tab:changes_in_education}
\end{table}

Prior research has extensively examined the impact of AI on computing education, particularly how students program with AI tools \cite{matinamoozadeh_2024_studentai, prather_2024_the, denny_2022_conversing, lucchetti_2024_substance, prather_2023_its}. Studies have shown that AI can be used to solve course assignments \cite{denny_2024_computing, denny_2022_conversing} and provide advanced tutoring and feedback mechanisms \cite{denny_2024_desirable, honolulu_2024_codeaid, liffiton_2023_codehelp}, which in turn necessitate changes in instructional design and assessment methods \cite{tankelevitch_2024_the}. A growing body of work agrees that Computer Science and Software Engineering education must redefine the core skills that educators emphasize \cite{lau_2023_from, annapurnavadaparty_2024_cs1llm}. To address this, we begin by discussing changes in individual development processes and then propose future directions for Computer Science and Software Engineering curricula, grounded in the industrial insights revealed by our data.

\subsection{Individual development process change}
In traditional programming practice, developers rely on just-in-time learning to fill knowledge gaps, such as searching for code examples, clarifying conceptual understanding, recalling syntax, or debugging implementation issues \cite{skripchuk_2023_analysis, chatterjee_2020_finding, brandt_2009_two}. Novice programmers often struggle to adapt online examples to their own programs, requiring repeated modification and experimentation before achieving a working solution \cite{ichinco_2015_exploring, wang_2021_novices}. But with LLMs, learners can bypass these barriers, which can hinder the critical element in programming - computational thinking skills \cite{wing_2006_computational}. 

\begin{figure}[hbt!]
  \centering
  \includegraphics[width=0.9\columnwidth]{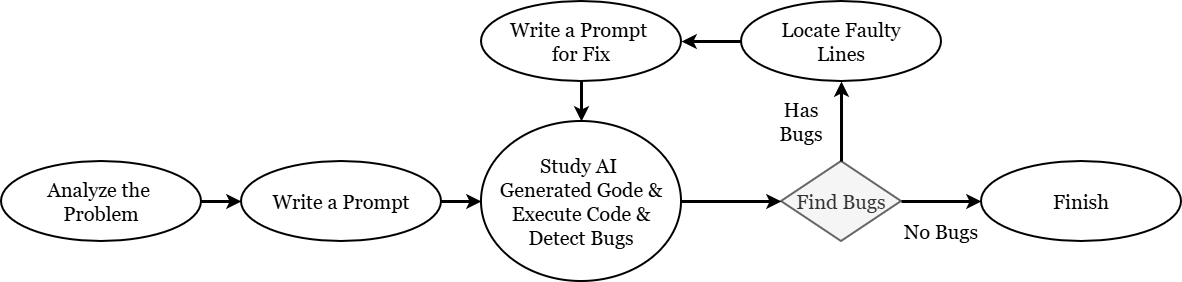}
  \caption{Vibe (Prompt-oriented) programming process}
  \label{fig:vibe_coding_process}
\end{figure}

\begin{figure}[hbt!]
  \centering
  \includegraphics[width=0.9\columnwidth]{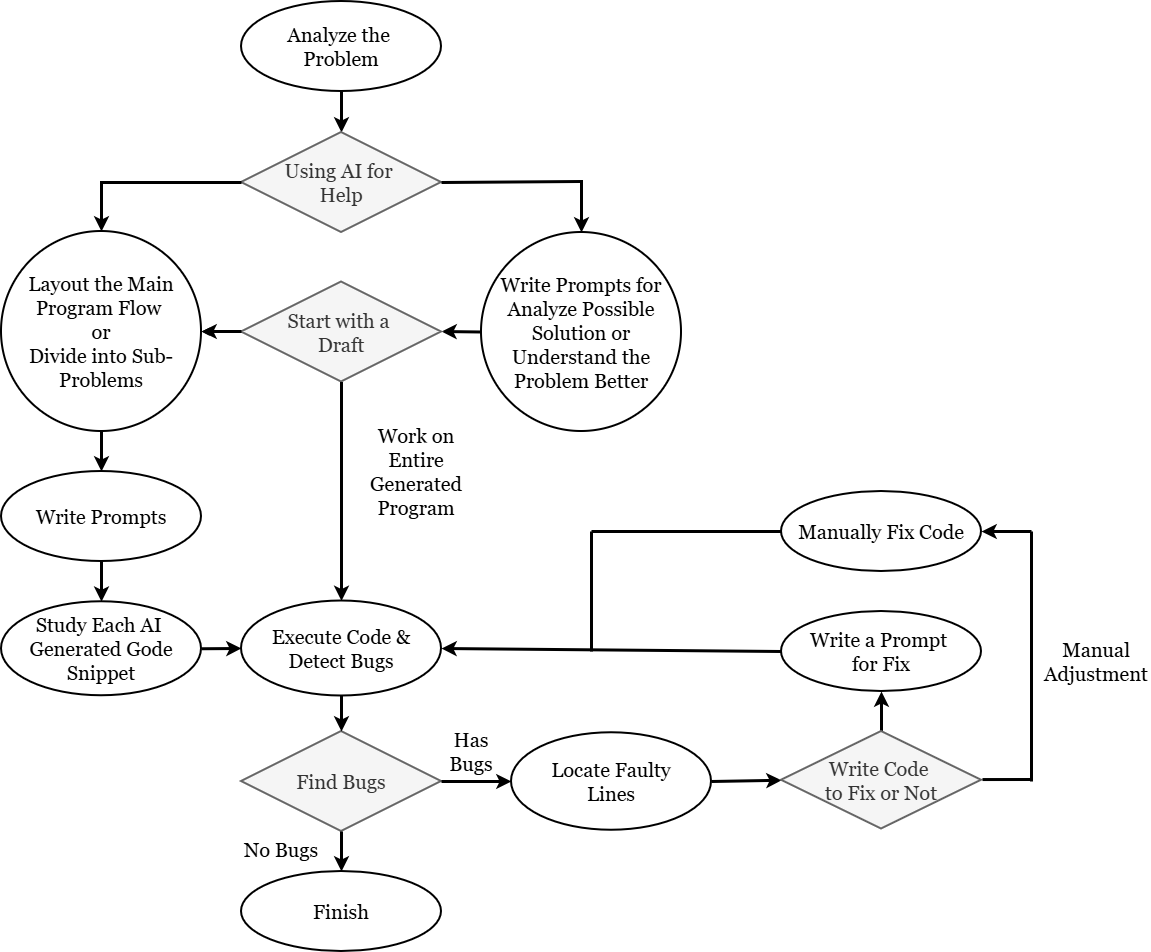}
  \caption{AI-assisted (hybrid) programming process}
  \label{fig:ai_assisted_coding_process}
\end{figure}

We investigate AI-based coding practices and identify two contrasting development processes: a prompt-oriented workflow (vibe coding) and an AI-assisted (hybrid) workflow (see \autoref{fig:vibe_coding_process} and \autoref{fig:ai_assisted_coding_process}). The AI-assisted workflow is informed by the procedural model proposed by Wang et al. \cite{wang_2023_the}.

These shifts introduce new forms of expertise and redirect how programming competence is developed. Traditional computer science education has historically emphasized the cultivation of formal implementation skills, including mastery of programming syntax, algorithmic reasoning, software frameworks, and the development of tacit system knowledge. Students are trained to internalize common idioms, mentally simulate program execution, and refine solutions through repeated coding practice.

With the introduction of LLM tools, expertise development increasingly emphasizes communication and orchestration rather than manual code production. Students are no longer positioned primarily as low-level code producers but as guides who communicate to AI systems. Key competencies in this paradigm include articulating precise prompts, critically evaluating AI-generated code, and planning development strategies.

Although prior studies suggest that novice learners may acquire knowledge through analyzing AI-generated code \cite{chow_2025_from, gadde_2025_democratizing}, significant concerns remain among educators and learners. Echoing industry practitioners’ worries, heavy reliance on AI-generated code may degrade programming expertise and computational thinking skills \cite{lau_2023_from, yilmaz_2023_augmented, zastudil_2023_generative}. Learners may develop a false sense of competence by accepting generated solutions without fully understanding or practicing them \cite{prather_2024_the}, which can negatively affect their ability to independently create, modify, or debug code without AI support \cite{bastani_2024_generative, majeedkazemitabaar_2023_how}. To mitigate these risks, educational practices must encourage deeper cognitive engagement with AI-generated artifacts.

\subsection{Toward Future Changes in Computer Science and Software Engineering Education}
The use of LLM tools in industrial software development is now unavoidable, and their presence in educational contexts is similarly inevitable. Based on insights derived from our data analysis, and in alignment with existing academic perspectives, we propose several directions for future Computer Science and Software Engineering education.

\subsubsection{Concentrate on Problem-solving}
Practitioners are concerned about the loss of problem-solving enjoyment and personal skill erosion underscore the continued centrality of problem-solving in software development. While LLM tools can reduce repetitive syntax-focused practice, they may also weaken individual problem-solving abilities. Prior studies suggest that reducing emphasis on syntax mastery can free instructional time to focus on algorithms and problem-solving \cite{majeedkazemitabaar_2025_exploring, denny_2024_computing, porter_2023_learn}. Educational programs can strengthen computational thinking and problem decomposition rather than language-specific syntax \cite{hutson_2023_rethinking}.

In this context, learners should be encouraged to actively evaluate and reason about AI-generated code, identify logical flaws, and debug logical errors. Instructional strategies may include requiring learners to explain generated solutions, justify design decisions, or reconstruct solutions without AI assistance. Instructors can employ an approach that involves learning through AI-generated examples with proper control on AI usage first, followed by many independent practices without AI assistance to reinforce understanding. Overall, curricula should prioritize problem framing, exploration of alternative solutions, and outcome evaluation, skills that remain essential in both traditional and AI-augmented development environments.

\subsubsection{Focus on Architectural Thinking and Specification-Driven Development}
Industry observations regarding role transformation suggest an increased need for architectural thinking in education. Architectural thinking is a design-oriented practice that concentrates on defining, analyzing, and reasoning about a system’s structure (i.e., components, interfaces, and configurations) to meet specific functional and quality requirements \cite{garlan_2000_software, bass_1998_software, fairbanks_2023_software}. Learners can be taught to practice holistic view on a software system to develop architectural thinking, which equips learners to reason about architecture, requirements, integration, emergent behavior, and organizational factors \cite{libardo_2023_training}. 

To support this capability, learners need to develop the ability to precisely describe the requirements and specifications. Specifications serve as detailed blueprints that define architecture, components, interfaces, and data flows, providing a foundation for implementation and iterative refinement. Writing precise specifications requires developers to adopt an architecture-level perspective, proficiency in describing system features and behaviors in a structured, precise, and unambiguous manner. Encouraging learners to engage with specification-based tasks can foster architecture-level reasoning and structured problem decomposition. Prior work has similarly emphasized the educational value of specification-based tasks \cite{spichkova_2022_refinementbased, suri_2002_introducing}.

This emphasis aligns with industrial discussions on prompt writing as a mechanism for expressing and validating system behavior when interacting with LLMs. Recent developments, such as GitHub’s Spec Kit, further reinforce this shift by treating specification writing as a foundational activity throughout the software development life-cycle.

\subsubsection{Introducing Project-based Learning Early}
Increasing importance of skills, such as code review, precise feature description, and new documentation formats, are exposed by practitioners. Project-based learning (PBL) is well suited to cultivating these competencies and has long been recognized as an effective approach for bridging theory and practice in computing education \cite{kolmos_2014_problembased, zhang_2024_from, saad_2022_the}. Rather than working on a well-scaffold assignment, doing a real-world project allows learners discover problems by themselves and integrate conceptual knowledge with hands-on development, fostering both technical competence and professional readiness. Students can also investigate and experience what practitioners mention about the shifted bottleneck in software development.

LLMs’ ability to generate code for small-scale tasks creates opportunities for students to engage earlier with large-scale, real-world systems. Through PBL, learners can practice system decomposition, code review, testing, integration, and deployment (i.e., core professional skills) while also developing teamwork, communication, and project management abilities. Given documented concerns about the accuracy and reliability of AI-generated code \cite{matinamoozadeh_2024_studentai, kazemitabaar_2023_studying, liang_2024_a}, project contexts provide a natural setting for teaching critical evaluation through review and testing. At the same time, learners might also gain insights into implementation strategies by analyzing AI-generated code \cite{chow_2025_from,gadde_2025_democratizing}. 

Furthermore, operational and infrastructural engineering practices like version control can also be exercised through a project. Modern software systems involve multiple frameworks, libraries, and development tools. Integrating LLMs into project-based learning enables learners to experience these technologies early in their education. 

Finally, project-based learning provides an effective means for fostering creativity, a goal long emphasized in computing education research \cite{groeneveld_2021_exploring, yadav_2017_fostering, salgian_2013_teaching}. LLM tools can rapidly support prototyping and brainstorming \cite{lim_2024_rapid, chang_2025_a}, enabling exploratory design and ideation. By reducing constraints imposed by syntax and low-level implementation details, these LLM tools allow learners to generate and evaluate ideas more efficiently. Moreover, AI can facilitate divergent thinking and experimentation, encouraging learners to explore multiple solution paths and design alternatives. This iterative exploration ultimately promotes creativity and supports the development of more innovative system designs.

\section{Limitation}
We acknowledge that this study has several limitations. First, the dataset is necessarily constrained by the limited time window of data collection and by reliance on publicly available YouTube content, which may not fully represent the diversity of industrial practices. In addition, aspects of the filtering, assessment, and interpretation processes are subject to the authors’ judgments, which may introduce potential bias despite cross-validation efforts.

Second, vibe and agentic coding practices, as well as LLM technologies themselves, are still rapidly evolving. The findings reported in this study might capture only a snapshot of an emerging phenomenon rather than a mature and stable practice. As tools, workflows, and professional norms continue to develop, practitioners’ experiences and perceptions are likely to change substantially. These limitations highlight the need for future research based on larger and more diverse datasets, as well as longitudinal and comparative studies, to examine how AI-based software development practices, required competencies, and tool ecosystems co-evolve over time.

\section{Conclusion}
This study systematically analyzes 57 YouTube video sources to uncover practitioners’ perspectives on the use of LLM tools in software development. Specifically, we examine how LLM tools are applied in practice, practitioners’ experiences with these tools, their concerns and perceptions of potential risks, and the resulting changes in development workflows. Based on the insights derived from our analysis, we further discuss the implications for computing education and highlight potential adaptations informed by the lessons learned.

Our findings reveal a fundamental shift in software development practices with the adoption of LLM-based tools. Practitioners describe a spectrum of AI-supported approaches, vibe coding, AI-assisted coding, and agentic coding, that differ in levels of AI autonomy and human involvement. Across these approaches, code generation is significantly accelerated, while development bottlenecks shift toward code review, testing, security assessment, and system-level reasoning. As a result, skills related to reading, evaluating, and validating code become increasingly critical, alongside heightened concerns about code quality, maintainability, traceability, and security risks introduced by AI-generated artifacts. At the same time, practitioners report notable changes in developer roles and skill development. While AI tools lower entry barriers and enable rapid prototyping, they also raise concerns about skill erosion, reduced problem-solving practice, and diminished learning opportunities for beginners. 

These changes have direct implications for computing education. Our analysis suggests that educational programs must adapt by shifting emphasis from syntax-centric training toward problem-solving, architectural thinking, and specification-driven development. Project-based learning emerges as a key pedagogical approach to expose students early to real-world workflows, including code review, security considerations, and collaboration with LLM tools, while also supporting creativity and innovation. Aligning curricula with these evolving industrial practices is essential to prepare future developers for AI-integrated software development environments.

\bibliographystyle{unsrt}  
\bibliography{references.bib}  

\appendix
\clearpage
\section*{Appendix: Content Source}

\begin{table}[h!]
\centering
\begin{tabular}{c p{4.5in} p{1.2in}}
\toprule
NO & Title and URL & Type \\
\midrule
1  & \raggedright Vibe Coding Is The WORST IDEA Of 2025 \\ \url{https://www.youtube.com/watch?v=1A6uPztchXk} & Opinion/Experience \\
2  & \raggedright Vibe coding, without the security Nightmares \\ \url{https://www.youtube.com/watch?v=olewFmJYEGI} & Discussion \\
3  & \raggedright Vibe Coding: Cybersecurity Risks and Security Best Practices \\ \url{https://www.youtube.com/watch?v=nT5y8ysy09Y} & Opinion/Experience \\
4  & \raggedright I Tried Vibe Coding — Here Are My Thoughts \\ \url{https://www.youtube.com/watch?v=bjh7EYdFTo4} & Opinion/Experience \\
5  & \raggedright I Tried Vibe Coding for 30 days — Lessons Learned \\ \url{https://www.youtube.com/watch?v=PDMxbbejgcA} & Opinion/Experience \\
6  & \raggedright The Best AI Coding Assistants | August 2025… interesting results \\ \url{https://www.youtube.com/watch?v=3C4TWUlkBMs} & Opinion/Experience \\
7  & \raggedright Vibe Coding Is The Future \\ \url{https://www.youtube.com/watch?v=IACHfKmZMr8} & Discussion \\
8  & \raggedright Context Engineering is the New Vibe Coding \\ \url{https://www.youtube.com/watch?v=Egeuql3Lrzg} & Opinion/Explanation \\
9  & \raggedright Blanca Champetier - Vibe Coding: A Game Changer for Experimentation and Finding Product Market Fit \\ \url{https://www.youtube.com/watch?v=JWvJUlssPqs} & Opinion/Experience \\
10 & \raggedright "I've changed my mind on AI coding" – Adam Wathan (creator of Tailwind) \\ \url{https://www.youtube.com/watch?v=X3yfVo2oxlE} & Discussion \\
11 & \raggedright Software Is Changing (Again) \\ \url{https://www.youtube.com/watch?v=LCEmiRjPEtQ} & Presentation \\
12 & \raggedright The “Vibe Coding” Mind Virus Explained \\ \url{https://www.youtube.com/watch?v=Tw18-4U7mts} & Opinion/Experience \\
13 & \raggedright What Is the "Vibe Coding" Mind Virus? \\ \url{https://www.youtube.com/watch?v=YlgmkCb5AdE&t=113s} & Opinion/Experience \\
14 & \raggedright Has This Report EXPOSED THE TRUTH About AI Assisted Software Development? \\ \url{https://www.youtube.com/watch?v=CoGO6s7bS3A} & Opinion/Experience \\
15 & \raggedright Is vibe coding even sustainable? \\ \url{https://www.youtube.com/watch?v=7WaMzm_PDpo} & Opinion/Experience \\
16 & \raggedright Vibe Coding: The No Code AI Revolution \\ \url{https://www.youtube.com/watch?v=Lb-OWjZE7q4} & Opinion/Experience \\
17 & \raggedright Who's Coding Now? - AI and the Future of Software Development \\ \url{https://www.youtube.com/watch?v=6Z5hlKIDV44} & Discussion \\
18 & \raggedright Vibe Coding Is a Problem \\ \url{https://www.youtube.com/watch?v=2b_KlROMfp8} & Discussion \\
19 & \raggedright Is Agent force the future of enterprise vibe coding? \\ \url{https://www.youtube.com/watch?v=Jw85_aD1xgY} & Discussion \\
20 & \raggedright Everything You Need to Know About Coding with AI // NOT vibe coding \\ \url{https://youtu.be/5fhcklZe-qE?si=2G8vY4FMGm0V7ScD} & Opinion/Experience \\
21 & \raggedright Vibe Coding at Scale: Customizing AI Assistants for Enterprise Environments \\ \url{https://www.youtube.com/watch?v=i1uPAN6uW4s} & Presentation \\
22 & \raggedright Vibe Coding Is The Future \\ \url{https://www.youtube.com/watch?v=riyh_CIshTs} & Discussion \\
23 & \raggedright Vibe Coding: Everything You Need to Know — With Amjad Masad \\ \url{https://www.youtube.com/watch?v=dCZc2kwtUIg} & Discussion \\
24 & \raggedright We Need to Talk About Vibe Coding — Thoughtworks panel? \\ \url{https://www.youtube.com/watch?v=Z-DoGQdlEPY} & Discussion \\
25 & \raggedright White Collar Jobs, Hyperscalers, AI Coding, Open vs Closed, Agents, and more! (Matt Garman) \\ \url{https://www.youtube.com/watch?v=nfocTxMzOP4} & Discussion \\
\bottomrule
\end{tabular}
\end{table}

\begin{table}[!ht]
\centering
\begin{tabular}{c p{4.5in} p{1.2in}}
\toprule
NO & Title and URL & Type \\
\midrule
26 & \raggedright The Future of Vibe Coding w/ Warp CEO Zach Lloyd (interview) \\ \url{https://www.youtube.com/watch?v=sn2wL_JlyjA} & Discussion \\
27 & \raggedright The Future of Software Development - Vibe Coding, Prompt Engineering \& AI Assistants \\ \url{https://www.youtube.com/watch?v=EIPxf7rgIPI} & Presentation \\
28 & \raggedright The "right way" to vibe code (engineers, please watch) \\ \url{https://www.youtube.com/watch?v=6TMPWvPG5GA} & Opinion/Experience \\
29 & \raggedright 16 Ways to Vibe Code Securely \\ \url{https://www.youtube.com/watch?v=0D9FMFyNBWo} & Opinion/Experience \\
30 & \raggedright When Vibe Coding, Avoid These 6 Security Risks! \\ \url{https://www.youtube.com/watch?v=5kWLPYAL8f0&t=336s} & Opinion/Experience \\
31 & \raggedright Amjad Masad: Vibe Coding, Platform Risk, Agentic Future, Permanent Underclass, and more! \\ \url{https://www.youtube.com/watch?v=s6TKlCdKbIs} & Discussion \\
32 & \raggedright Enterprise-Grade AI: A Vibe Coding Demo (enterprise demo/report) \\ \url{https://www.youtube.com/watch?v=h2iriY5VFZ0} & Presentation \\
33 & \raggedright Vibe Coding in Production: A Founder/CTO’s 2025 AI Engineering Playbook \\ \url{https://www.youtube.com/watch?v=wW_nseHqalg} & Presentation \\
34 & \raggedright Dave Plummer — Vibe coding and the future of programming \\ \url{https://www.youtube.com/watch?v=f15bhW7juwI} & Discussion \\
35 & \raggedright Vibe Coding \& AI Agents — The Future of Software \\ \url{https://www.youtube.com/watch?v=YxUZIZs4JKk} & Discussion \\
36 & \raggedright AI’s Vibe-Coding Era — How The Shift To Apps Changed The Race \\ \url{https://www.youtube.com/watch?v=mmws6Oqtq9o} & Discussion \\
37 & \raggedright Cognizant Vibe Coding Week CEO Conversations | Cognizant x Lovable \\ \url{https://www.youtube.com/watch?v=dXlBLvJU1bE} & Discussion \\
38 & \raggedright AppSec for the Vibe Coding Era — GitHub AppSec AI Summit \\ \url{https://www.youtube.com/watch?v=wgTNqYkoICM} & Presentation \\
39 & \raggedright How to Build Production-ready Applications with Vibe Coding \\ \url{https://www.youtube.com/watch?v=vGm3WMUUOHU} & Discussion \\
40 & \raggedright From Vibe to Production: Taming AI for Fearless Development \\ \url{https://www.youtube.com/watch?v=8tFjsJBhA6k} & Presentation \\
41 & \raggedright Vibe Coding What Could Possibly Go Wrong \\ \url{https://www.youtube.com/watch?v=YWX3xjyJGyU} & Discussion \\
42 & \raggedright Vibe Coding is Here - How AI is Changing How We Build Online \\ \url{https://www.youtube.com/watch?v=xxA-M3HrKrc} & Discussion \\
43 & \raggedright How vibe coding can destroy your project \\ \url{https://www.youtube.com/watch?v=vaKfVgGk3EY} & Opinion/Experience \\
44 & \raggedright Busting the AI Coding Productivity Myth \\ \url{https://www.youtube.com/watch?v=MLSs8Asmthk} & Opinion/Experience \\
45 & \raggedright Future of vibe coding | DHH and Lex Fridman \\ \url{https://youtu.be/wz65rRHL6jM?si=OBpHxhZ9aqGSUVtp} & Discussion \\
46 & \raggedright Here is Why Vibe Coding is a Dead End for Juniors and Non-programmers \\ \url{https://youtu.be/fzvx2bEUUnY?si=vsm9T6g-xSBne95y} & Opinion/Experience \\
47 & \raggedright Is Vibe Coding Dead? Here’s What you Need to Know About Tools Like Lovable \\ \url{https://youtu.be/cJdVasCOOPE?si=82KqxbzbMfME1Jcj} & Opinion/Experience \\
48 & \raggedright The Rise And Fall Of Vibe Coding: The Reality Of AI Slop \\ \url{https://youtu.be/vHPpBZiR80c?si=-aAz0TszxObPMWtt} & Opinion/Experience \\
49 & \raggedright Vibe Coding 101 for Non-Coders: How Small Teams Can Build Real Tools with AI \\ \url{https://youtu.be/j-H0Na0cHTs?si=4gJyBOyQTyfwvE_B} & Discussion \\
50 & \raggedright Vibe Coding vs Low-Code/No-Code: Security Risks and CI/CD Pipeline Impacts for Citizen Developers \\ \url{https://youtu.be/VCWQoLQ9PhM?si=w3yUyNVdQSOjkwjy} & Discussion \\
\bottomrule
\end{tabular}
\end{table}

\clearpage


\begin{table}[h!]
\centering
\begin{tabular}{c p{4.5in} p{1.2in}}
\toprule
NO & Title and URL & Type \\
\midrule
51 & \raggedright Vibes won't cut it — Chris Kelly, Augment Code \\ \url{https://youtu.be/Dc3qOA9WOnE?si=BKML_q2LuRsnzbKk} & Presentation \\
52 & \raggedright Interview with King of AI Coding (Replit CEO - \$1.2B) \\ \url{https://youtu.be/r45-w5MKOlE?si=2JtEBa6oTRM8n9cX} & Discussion \\
53 & \raggedright Replit CEO on The Career of Coding, AGI, and Vibe Coding Wars w/ Amjad Masad, Dave B \& Salim \\ \url{https://youtu.be/T9rukIoXxgw?si=W1SJ01MqxZtwVWIC} & Discussion \\
54 & \raggedright Vibe coding is already dead \\ \url{https://youtu.be/tKPtZtsLgUA?si=5_14NsZ1qjwp3OQv} & Opinion/Experience \\
55 & \raggedright Vibe coding or AI pair? Best practices for using AI coding assistants \\ \url{https://youtu.be/7ZOcPiEgPvc?si=KtQSBOJdGTFUJCQR} & Discussion \\
56 & \raggedright Why "Vibe Coding" Fails in Enterprise: AI's Real Limits Exposed \\ \url{https://youtu.be/q61UFd2mkpw?si=u4EFBc_-AwlvJaim} & Discussion \\
57 & \raggedright Startup Founder Best Practices - AI Meetups and Vibe Coding w/Ben Cheatham \\ \url{https://www.youtube.com/live/NGZc0T3CG3M?si=YTIKAWihBpyuyqEy} & Discussion \\
\bottomrule
\end{tabular}
\end{table}

\end{document}